# Coulomb-Blockade Transport in Quasi-One Dimensional Polymer Nanofibers

A. N. Aleshin[1,2,*], H. J. Lee[1], S. H. Jhang[1], H. S. Kim[1], K. Akagi[3], and Y. W. Park[1]

[1]*School of Physics and Nano Systems Institute - National Core Research Center,
Seoul National University, Seoul 151-747, Korea*
[2]*A. F. Ioffe Physical-Technical Institute, Russian Academy of Sciences, St. Petersburg 194021, Russia*
[3]*Institute of Materials Science and Tsukuba Research Center for Interdisciplinary Materials Science,
University of Tsukuba, Tsukuba, Ibaraki 305-8573, Japan*

We report the low temperature current-voltage (I-V) characteristics studies in quasi one-dimensional conducting polymer nanofibers. We find a threshold voltage $V_t$ below which little current flows at temperatures below 30 - 40 K. For $V > V_t$ current scales as $(V/V_t - 1)^\zeta$, where $\zeta \sim 1.8 - 2.1$ at high biases. Differential conductance oscillations are observed whose magnitude increases as temperature decreases below 10 K. We attributed the observed low temperature I-V behavior to Coulomb blockade effects with a crossover to Luttinger liquid- like behavior at high temperature. We demonstrate that at low temperatures such a doped conjugated polymer fiber can be considered as an array of small conducting regions separated by nanoscale barriers, where the Coulomb blockade tunneling is the dominant transport mechanism.

Complex structures built of nanosize conducting objects provide model systems for investigation of transport phenomena on the mesoscopic scale, where quantum confinement and Coulomb charging play an important role [1]. Electronic transport through an array of metallic nanocrystals separated by nanobarriers is determined by the interplay between single-electron charging of an individual conducting region and tunneling between adjacent islands. In such systems both the tunneling resistance between neighboring regions is large, $R_T \gg h/e^2$ and the charging energy $E_C = e^2/2C$ of an excess electron on a site is larger than $k_B T$ (C is the capacitance of the region). In the presence of both charge and structural disorder this interplay leads to highly non-Ohmic current-voltage (I-V) characteristics. Coulomb blockade theory predicts that at low temperatures, there is no current below a specific threshold voltage, $V_t$, while above $V_t$ the current follows a power law:

$$I \sim (V - V_t)^\zeta \qquad (1)$$

with $\zeta \sim 1$ in one dimensional (1D) and 5/3 or 2 in two-dimensional (2D) systems [2]. $V_t$ depends on the nanocrystal number, the capacitance of the conducting regions and the capacitance between the each region and the back gate. Such a behavior with scaling exponents between 1 and 2.3 has been found, for example, in narrow chains of carbon nanoparticles [3], and in 100 nm wide multilayer of gold particles which exhibited $\zeta \sim 1.6$ [4]. Coulomb-blockade effects have also been observed in multiwalled carbon nanotubes [5], organic thin-film transistors based on highly ordered molecular materials [6] as well as in single-molecular transistor structures [7]. It is evident that the nature of the individual regions: metallic, semiconducting, quantum dots - is irrelevant for this phenomenon to be observed [8]. In view of these results the question arises whether Coulomb-blockade effects can also be observed in such inherent quasi-1D systems as conducting polymer nanofibers. In this connection polyacetylene (PA) nanofibers are of particular interest as the model system for nanotransport studies because of the simple chemical structure, well defined polycrystallinity [9], and good ability for doping [10-12]. It is shown that at temperature 30 K < T < 300 K doped polymer nanofibers demonstrate some transport features characteristic of quasi-1D systems, namely, series of Luttinger liquids separated by intra molecular barriers [13]. This implies the power-law behavior in both the temperature dependence of the conductance, G(T), and I-V characteristics, I(V) [14]. However, this approach does not work at temperature below 30 K, where I-Vs of doped polymer nanofibers revealed strongly non-Ohmic behavior [15,16]. At present the complete understanding of the low temperature transport mechanism including the Coulomb-blockade tunneling in a single conducting polymer fiber is still lacking. In this paper we demonstrate that the low temperature I-V behavior of quasi-1D polymer fibers can be attributed to Coulomb blockade effects with a crossover to Luttinger liquid-like behavior at high temperature. We show that at low temperatures such a doped polymer fiber can be considered as an array of small conducting regions separated by nanoscale barriers where the Coulomb blockade tunneling is the dominant transport mechanism.

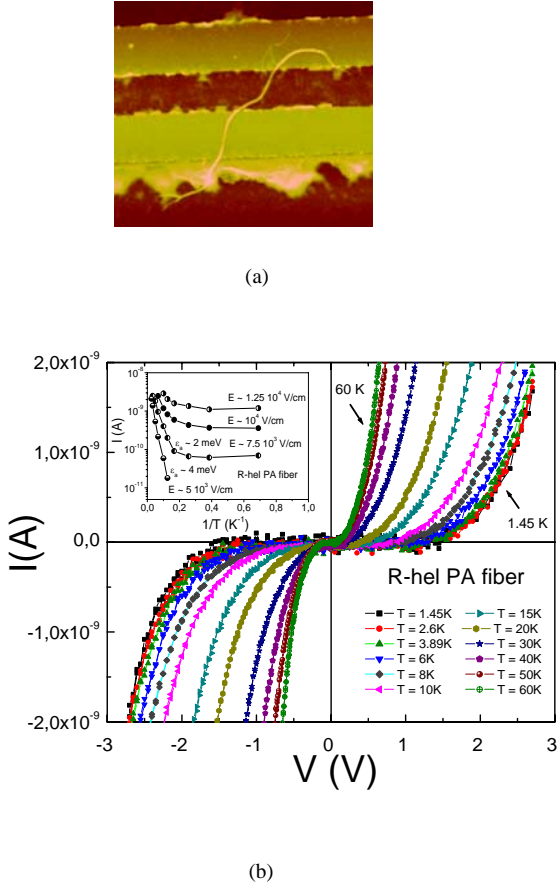

(a)

(b)

Fig. 1. (a) AFM image of R-hel PA fiber used in our study on top of Pt electrodes. The fiber cross-section is ~ 42 nm (high) and ~ 220 nm (wide). (b) Typical I-V characteristics of R-hel PA fiber at low temperatures. Inset: temperature dependence of current at different electric fields for the same sample.

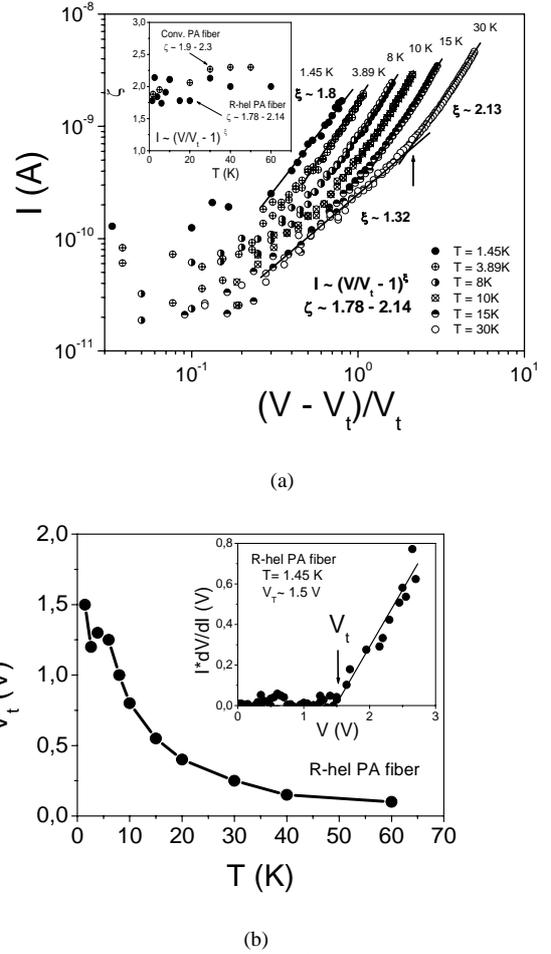

(a)

(b)

Fig. 2. ((a) Current vs. $(V - V_t)/V_t$ at different temperatures. Inset: power exponent, $\zeta$ vs. T for the R-hel PA fiber and for the conventional iodine doped PA fiber (diameter 16 nm) estimated from data reported in Ref. 15. (b) Temperature dependence of the threshold voltage. Inset shows the IdV/dI vs. V and the threshold voltage, $V_t$ ~ 1.5 V, at T = 1.45 K for the same R-hel PA fiber.

We have studied helical PA (hel PA) nanofibers synthesized using chiral nematic liquid crystal as a solvent of Ziegler-Natta catalyst following the procedure presented elsewhere [17]. Hel PA has polycrystalline structure [18] and can be made with either R- (counterclockwise) or S- (clockwise) type helicity. Single R-hel PA fibers with a cross-section, typically 40 - 60 nm (high), 100 - 300 nm (wide) and the length up to ~ 10 μm, were deposited on a Si substrate with a SiO$_2$ layer and platinum electrodes thermally evaporated on the top, 2 μm apart. The fiber was doped with iodine from vapor phase up to the saturation level similar to our previous studies [13, 19]. The transport measurements were performed in the 2-probe geometry in a vacuum at ~$10^{-5}$ Torr, using Janis cryostat with a Keithley 6517A electrometer.

Figure 1(a) shows the AFM image of typical R-hel PA nanofiber used in our study on top of Pt electrodes. Figure 1(b) presents the low temperature I-V characteristics for such R-hel PA nanofiber. As can be seen from Fig. 1(b), the I-Vs at T < 60 K are symmetric and strongly non-linear with a narrow Ohmic region even at relatively high temperatures. Inset to Figure 1(b) shows that the temperature dependence of a current, I(T), has an activated nature with small activation energies, $\varepsilon_a$ ~ 2 - 4 meV at high temperatures and low electric fields, while at lower temperatures and at high electric fields the I-Vs are almost temperature independent. It is worthy to note, that the similar I(T) behavior was found earlier in such quasi-1D systems as conventional PA nanofibers [15,16], polydiacetylene crystals and films [20,21]. As can be seen from Figs. 1(b) and 2(a), as temperature goes below 30 - 40 K, the current starts to flow only above some voltage threshold, $V_t$, with negative temperature coefficient of $V_t$ ($V_t$ decreases as temperature increases, as shown in Fig. 2(b)). There is a pronounced power-law behavior in G(T) and I(V) at T > 30 – 40 K for the same R-hel PA fiber with a good I-Vs scaling as $I/T^{\alpha+1}$ versus $eV/k_BT$ (Fig. 3(a)), where $\alpha$ ~ 4.4 is the exponent in G(T) ~ $T^{\alpha}$, similar to our previous report [13]. As temperature decreases, the I-Vs of R-hel PA nanofiber still follow the power law, but the G(T) plot is not available due to presence of the $V_t$. As can be seen from Fig. 2(a), the I-Vs at low temperature, $T < 30 - 40$ K, and high biases, $eV \gg k_BT$, follow another power-law scaling fitted by Eq. (1), with nonzero $V_t$ which can be estimated from $IdV/dI$ vs. $V$ plots at each temperature. Figure 2(b) and inset demonstrate that $V_t \sim 1.5$ V at $T = 1.45$ K and $V_t$ decreases below this value with increasing temperature. The scaling exponents, $\zeta$, obtained from $I$ vs. $(V - V_t)/V_t$ log-log plots at high biases are found to be $\zeta \sim 1.78 - 2.14$ and weakly temperature dependent (Fig. 2(a) and inset to Fig. 2(a)). The same behavior has been found for several other iodine doped R-hel PA nanofibers studied in this work. There are some features in low temperature I-Vs of doped R-hel PA nanofibers. First, as can be seen from Fig. 2(b), at $T > 4$ K the temperature dependence of the $V_t$ rather follows the power law $V_t \sim T^{-\gamma}$ with the power exponent $\gamma \sim 1$. Secondly, there is an apparent transition in $I$ vs $(V - V_t)/V_t$ plots at $T > 30$ K from the scaling exponent, $\zeta \sim 1.3$ at low biases to $\zeta \sim 2.1$ at higher biases, as shown in Fig. 2(a). The cusp point shifts to the lower bias values as temperature decreases, thus at the lowest temperature, $T \sim 1.45$ K, we observed the only slope with $\zeta \sim 1.8$. Inset to Fig. 3 (a) shows the differential conductance $(dI/dV)$ derived from the I-V characteristics of the R-hel PA nanofiber at different temperatures. The $dI/dV$ vs $V$ plots demonstrate a clear parabolic behavior of the conductance at high temperature with a pronounced conductance oscillations whose magnitude greatly increases with decreasing temperature. Despite of the presence of random noise contribution, one can estimate the period of these oscillations at the lowest temperature $T = 1.45$ K as $\sim 200 - 300$ mV.

It is worthy to note that the low temperature transport features of conventional PA nanofibers have been discussed earlier [15,16] in the framework of the Zener tunneling model [22]. However, as was shown in Ref. 21, the probability of Zener tunneling is lower than that for a charge injection to the polymer fiber from the metallic banks. This injection is provided by tunneling of the electrons (holes) into the conduction (valence) bands of the polymer. Such a tunneling is possible since, due to a presence of high electric field within the polymer fiber, the corresponding band edges have finite slopes and at some distance from the bank, the band edge coincides with the Fermi level within the bank. One also expects that disorder induces some scatter between the positions of the band edges within the different parts of nanofibers of the order $\partial\varepsilon$. In this case the observed activation energy $\varepsilon_a \sim 2 - 4$ meV might be related for some extent to the band gap mismatch between the different conductive islands and polymer nanofibers [21].

From another hand the characteristic low temperature I-V behavior of doped R-hel PA nanofibers is similar to that observed in 2D metal nanocrystal arrays [2,4,23,24] as well as in 1D chains of graphitized carbon nanoparticles [3], where the Coulomb blockade is considered as the dominant transport mechanism. In the case of iodine doped R-hel PA nanofibers the localization of electron states by disorder at low temperature results in the appearance of the transport threshold voltage - $V_t$. The existence of $V_t$, its temperature dependence, the scaling exponents $\zeta \sim 1.78 - 2.14$ in Eq. (1) are characteristic of the Coulomb blockade transport similar to that in quasi-1D arrays of metallic nanoparticles, separated by tunneling nanobarriers [3,23,24]. The general parabolic shape of the differential conductance at high temperature and clear low temperature conductance oscillations indicates tunneling through the entire metal-insulator-metal junction for a structure with multiple tunnel barriers in the Coulomb blockade regime. The results of our analysis of early data for conventional iodine doped PA fiber (diameter $\sim 16$ nm), available from Refs. 15,16, revealed the qualitatively similar Coulomb blockade behavior of current at $eV \gg k_BT$ and $T < 50$ K with a weakly temperature dependent power exponents $\zeta \sim 1.9 - 2.3$ in Eq (1) (see inset to Fig. 2(a)). Based on these experimental results we suggest that, at low temperature the quasi-1D polymer fiber can, to first order, also be considered as an array of small conducting regions (quantum dots) made of doped conjugated polymer separated by nanoscale barriers as shown in Fig. 3 (b). This suggestion is supported by results of X-ray diffraction studies which revealed the polycrystalline structure of R-hel PA [18]. Recent SEM studies have demonstrated that each R-hel PA nanofiber is composed of smaller helical nanowires that are bundles of polymer chains [25]. This structure implies the superposition of several conducting channels inside the main R-hel PA fiber and may affect the transport by leading to some scattering in power exponents, threshold voltages, and conductance oscillation periods. The Coulomb blockade transport in such an array of tunnel junctions implies that the I-Vs behavior depends on the size, dimension and connectivity of the arrays, the screening length and the disorder [8]. In the case of doped polymer nanofiber the size of an array of metallic regions separated by tunnel nanojunctions should be comparable to the size of polycrystalline grains and thus to the diameter of a smallest nanowire ~ some tens of nm. From another hand at the modest doping level of the PA chain the distance between impurities (iodine) is of the order 10 nm. Iodine atoms affect transport in the PA chain by the creation of conjugation defects, domain walls (soliton kinks), etc., which results in the creation of conductive islands of the size comparable to the distance between impurities. It is worthy of note that the Coulomb blockade transport in chains of graphitized carbon nanoparticles of diameter ~ 30 nm reported in Ref. 3 revealed the values of $V_t$ and $\zeta$ similar to those in case of R-hel PA fibers. All the above-mentioned arguments allow us to estimate the average size of conductive islands inside the R-hel PA nanofiber of the order ~ 10 nm. In the Coulomb blockade model the threshold voltage at zero temperature, $V_t(0)$ can be expressed as $V_t(0) = \gamma N V_0$, where N is the number of nanoparticles along the length of the array and $E_C = eV_0$ is the charging energy associated with single electron tunneling between neighboring particles in the array. For 2D close-packed nanoparticles arrays theory predicts $\gamma = 0.226$, while $\gamma = 0.5$ for 1D chains [2,23]. For quasi-1D gold nanoarrays the values of $\gamma \sim 0.34$ - between the 1D and 2D theoretical predictions - were found [26]. Taking into account the above-proposed average size of the conducting islands in the doped polymer fiber ~ 10 nm, and the distance between electrodes ~ 2 μm, one can estimate the approximate amount of polymer quantum dots in the polymer array as ~ 200. Therefore the charging energy $eV_0$ obtained using the parameter $\gamma \sim 0.34$ for quasi-1D arrays varies from ~ 22 meV at T = 1.45 K down to ~ 4 meV at T = 30 K. The latter value of the charging energy coincides well with an activation energy, $\varepsilon_a \sim 4$ meV, estimated from I vs 1/T plot for R-hel PA fiber at low electric fields and high temperature (inset to Fig. 1(b)) and argues in favor the Coulomb blockade model.

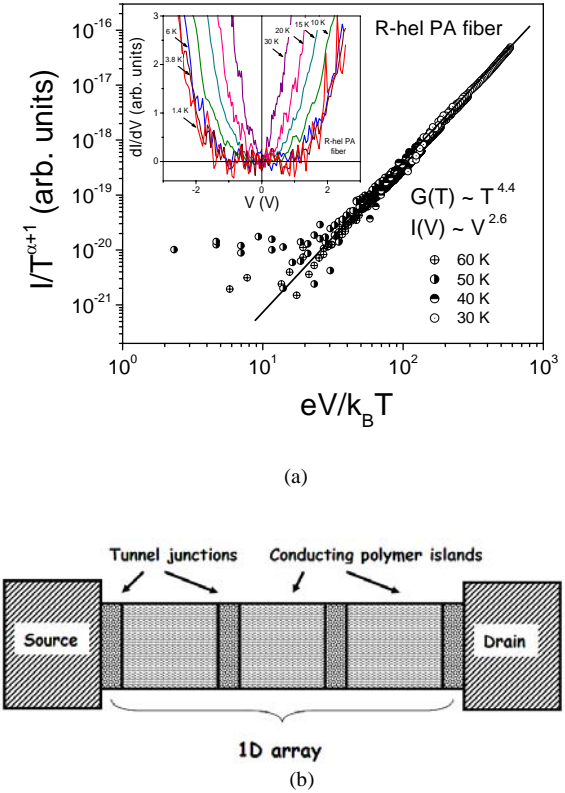

Fig. 3. (a) $I/T^{\alpha+1}$ vs. $eV/k_BT$ for R-hel PA fiber at $T > 30$ K; $\alpha \sim 4.4$ is the exponent in $G(T) \sim T^\alpha$ at high temperature. Inset shows differential conductance vs. voltage at different temperatures for the same R-hel PA fiber. Note the clear conductance oscillations whose magnitude increases with decreasing temperature. (b) Schematic of the metal-insulator-metal tunneling junction structure of a conducting polymer nanofiber.

The Coulomb blockade approach predicts that the scaling exponent $\zeta \sim 1$ is characteristic of 1D system, while $\zeta \sim 2$ is typical for 2D tunneling [2]. The values of scaling exponents $\zeta \sim 1.8 – 2.1$ obtained from the I vs. $(V – V_t)/V_t$ plots for R-hel PA fibers (at low temperature and at high biases) are characteristic of 2D tunneling in the close-packed nanoarray of conducting polymer islands (dots). The existence of two slopes at T > 30 K: namely $\zeta \sim 1.3$ at low biases and $\zeta \sim 2.1$ at high biases, can be attributed to the transition from the 2D tunneling in polymer nanoarray at low temperature to the quasi-1D Luttinger liquid-like transport at higher temperature, described in detail in Ref. 13, where $V_t \sim 0$ and thus the Ohmic regime in I-Vs becomes available. This behavior correlates well with the crossover from Luttinger liquid to Coulomb blockade regime observed in MWNT recently [27,28].

In conclusion, we found that the low temperature electronic transport features in quasi-1D conducting polymer nanofiber can be attributed to Coulomb blockade effects. We demonstrate that at low temperatures such a doped polymer fiber can be considered as an array of small conducting regions (quantum dots) separated by nanoscale tunneling junctions where the Coulomb blockade tunneling is the dominant transport mechanism. We suppose that such a behavior is a general feature of the low temperature transport in other types of conducting polymer nanofibers, which is an essential issue for potential practical applications in nanoelectronics.

This work was supported by the Nano Systems Institute - National Core Research Center (NSI-NCRC) program of KOSEF, Korea. Support from the Brain Pool Program of Korean Federation of Science and Technology Societies for A. N. A. is gratefully acknowledged.